\pdfoutput=1
\documentclass{IEEEtaes}
\usepackage{amsmath,amssymb,bm,graphicx,booktabs,url,cite}
\makeatletter
\def\figcapfont{\fontfamily{\rmdefault}\fontsize{8.25}{10}\selectfont}
\def\tablecapnumfont{\fontfamily{\rmdefault}\fontsize{8.25}{10}\selectfont\leftskip0pt\rightskip0pt plus1fill}
\def\tablecapfont{\fontfamily{\rmdefault}\fontsize{8.25}{10}\selectfont\leftskip0pt\rightskip0pt\parfillskip0pt plus1fil}
\long\def\@makecaption#1#2{\ifx\@captype\@IEEEtablestring{\tablecapnumfont#1\par}{\tablecapfont#2\par}\@IEEEtablecaptionsepspace\else\@IEEEfigurecaptionsepspace\vspace*{2pt}{{\figcapnumfont#1.}\hspace*{9.5pt}{\figcapfont#2\par}}\fi}
\def\p@subsection{\thesection-}
\makeatother
\jvol{XX}
\jnum{XX}
\jmonth{XXXXX}
\paper{1234567}
\pubyear{2026}
\doiinfo{TAES.2026.Doi Number}
\newtheorem{theorem}{Theorem}
\begin{document}

\title{A Cubic Cram\'er--Rao Path-Length Law for Scalar-Map Localisation}
\author{JOSEPH O'LEARY}
\affil{Swinburne University of Technology, Melbourne, Australia}
\author{BILL MORAN}
\affil{University of Melbourne, Melbourne, Australia}
\author{WENCHAO LI}
\affil{RMIT University, Melbourne, Australia}
\author{ANDREW D. GREENTREE}
\affil{RMIT University, Melbourne, Australia}
\author{ALLISON KEALY}
\affil{Swinburne University of Technology, Melbourne, Australia}
\receiveddate{Manuscript received XXXXX 00, 0000; revised XXXXX 00, 0000; accepted XXXXX 00, 0000.}
\corresp{{\itshape (Corresponding author: J. O'Leary.)}}
\authoraddress{J. O'Leary is with the Innovative Planet Research Institute, Swinburne University of Technology, Melbourne, Australia (e-mail: \href{mailto:jdoleary@swin.edu.au}{jdoleary@swin.edu.au}). B. Moran is with the Department of Electrical and Electronic Engineering, University of Melbourne, Melbourne, Australia. W. Li and A. D. Greentree are with the Department of Physics, School of Science, RMIT University, Melbourne, Australia. A. Kealy is with the Innovative Planet Research Institute, Swinburne University of Technology, Melbourne, Australia.}
\markboth{O'LEARY ET AL.}{CUBIC CRAM\'ER--RAO PATH-LENGTH LAW FOR SCALAR-MAP LOCALISATION}
\maketitle

\begin{abstract}
Map-referenced navigation compares onboard measurements of a terrain,
gravity, or magnetic anomaly field with a presurveyed map.
A scalar measurement constrains position along the local map gradient,
and two-dimensional position becomes identifiable only when the gradient
axis turns along the path.
Map suitability for navigation is commonly assessed by gradient
magnitude.
Here we prove that, on a short path of length $L$ with a constant
horizontal position error, the Fisher information for the least-informed
position component grows at most as $\kappa L^3$, where
$\kappa\propto(\|\nabla B\|\,\|\nabla\theta\|)^2$ is set by the gradient
magnitude $\|\nabla B\|$ and the turning rate $\|\nabla\theta\|$ of the
gradient angle $\theta$ at the start of the path.
Across 20,894 and 43,283 locations on two aeromagnetic anomaly maps, we
compute the shortest straight-path length $\ell_{\rm samp}$ at which the
Cram\'er--Rao lower bound (CRLB) on the least-informed position
component reaches 100~m.
Without an exhaustive search, $\kappa$ ranks locations by
$\ell_{\rm samp}$ with Spearman correlation $|\rho_{\rm S}|\geq0.96$, to
be compared with $|\rho_{\rm S}|\leq0.56$ for $\|\nabla B\|$.
The distance predicted from $\kappa$ lies within a factor 1.5 of
$\ell_{\rm samp}$ at $\geq85\%$ of locations, against $\approx50\%$ for
distances fitted to $\|\nabla B\|$.
In simulation, a maximum-likelihood estimator of the position error
reaches 100~m on paths with a median excess of $\leq11\%$ over
$\ell_{\rm samp}$, and hence nearly attains the CRLB.
Before a mission, the short-path law allows a navigator to (i) screen
map regions for localisation, (ii) choose the initial heading along
$\nabla\theta$, and (iii) relate sensor noise to the distance needed for
a target accuracy.
\end{abstract}


\section{INTRODUCTION}
\label{sec:introduction}

Positioning, navigation, and timing services, e.g.\ precision agriculture, mining, surveying, aviation, and maritime operations, rely on global navigation satellite systems [GNSS; \cite{Psiaki_2016}]. GNSS positioning is vulnerable or unavailable in at least two situations. First, GNSS signals received at Earth are weak~\cite{ISGPS_2004}, and hence are susceptible to jamming and spoofing~\cite{Psiaki_2016,Borio_2016}. Second, GNSS signals are attenuated underwater~\cite{Fernandes_2023}, underground~\cite{Zare_2021}, as well as inside buildings~\cite{Ruehrup_2020}, and are subject to multipath effects in urban canyons~\cite{Xu_2024}. An inertial navigation system~\cite{Woodman_2007} can bridge GNSS unavailability in either situation, but its position error grows with time, e.g.\ quadratically for a constant accelerometer bias~\cite{Woodman_2007}. Map-referenced navigation limits the foregoing growth by comparing onboard measurements with a presurveyed map of a geophysical field.

Map-referenced navigation requires a map of a spatially varying field, surveyed in advance, e.g.\ by spaceborne radar interferometry~\cite{Farr_2007}, or by terrestrial, airborne, and satellite gravimetry and magnetometry~\cite{Pavlis_2012,Meyer_2017}. Current navigation systems exploit terrain elevation~\cite{Golden_1980,Bergman_1999,Nordlund_2009,Carroll_2021}, the gravity anomaly~\cite{Wang_2017,Liu_2019,Li_2024}, and the magnetic anomaly field [MagNav; \cite{Canciani_2016,Canciani_2017,Lee_2020,OLeary_2026}]. Map-referenced navigation has been demonstrated at sea, in flight, and on foot outdoors and indoors, with reported errors from a few metres to a few kilometres~\cite{Salavasidis_2021,Everitt_2026,Muradoglu_2025,Scholes_2026,Solin_2016}, and magnetometers for MagNav have been operated on moving road vehicles~\cite{Graham_2025}.

\begin{figure}[t]
    \centering
    \includegraphics[width=\columnwidth]{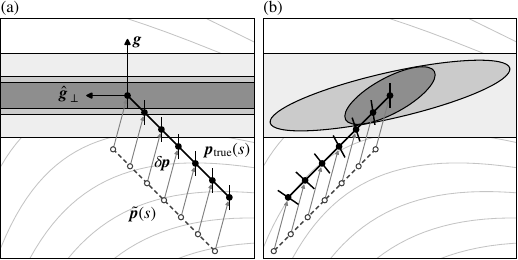}
    \caption{Cram\'er--Rao regions for a rigid shift $\delta\bm p$ along two straight tracks from the same start on a schematic scalar map $B$ (grey contours). The inertial track $\tilde{\bm p}(s)$ (dashed) is displaced from the true track $\bm p_{\rm true}(s)$ (solid) by $\delta\bm p$ (arrows). Shading shows the $1\sigma$ region for $\delta\bm p$ after one, four, and seven measurements, drawn at the start. (a)~The gradients $\bm g$ are parallel, and the region remains a strip along the contour tangent $\hat{\bm g}_\perp$. (b)~The gradient axis (short segments) turns, and the region closes. The turn is exaggerated for visibility. That is, a scalar map fixes the position in two dimensions only where the gradient axis turns along the path. The coefficient $\kappa$ of Section~\ref{sec:cubic} sets how quickly the region closes as the path lengthens.}
    \label{fig1}
\end{figure}

Onboard measurements of the mapped field can be scalar, e.g.\ terrain elevation, the gravity anomaly, or the magnetic field intensity, or vector- or tensor-valued, e.g.\ the magnetic field vector or the gravity gradient tensor. To first order, a scalar measurement constrains displacement along the local gradient and leaves displacement along the contour unconstrained, as in Fig.~\ref{fig1}(a). Two-dimensional position becomes identifiable only when the gradients sampled along the path are not all collinear, as in Fig.~\ref{fig1}(b). The Cram\'er--Rao lower bound [CRLB; \cite{Kay_1993}] quantifies how well a given path determines position. It has been evaluated for terrain maps as a function of the noise levels and the terrain gradient~\cite{Bergman_1999}, and for magnetic anomaly maps in terms of the spatial derivatives of the field~\cite{Tkhorenko_2022} and recursively with a Gaussian-process map~\cite{Siebler_2022}. Over a regional anomaly map, the CRLB tracks the anomaly gradient~\cite{Gupta_2024}, and the resulting horizontal error is anisotropic~\cite{Sengupta_2025}.

There are at least three separate challenges associated with map-referenced navigation. First, the map is an imperfect model of the field, with finite resolution and survey error, and the achievable accuracy depends on map quality~\cite{Gupta_2024}. Second, the onboard measurement contains contributions other than the mapped field, which must be removed or estimated, e.g.\ the magnetic field of the aircraft, which can be calibrated in flight~\cite{Canciani_2022,Hager_2026}. Third, the information a path collects depends on the map and on the whole path, so the distance and direction a vehicle must travel are found by search over headings and path lengths at every map location. Map-suitability measures built on gradient magnitude, e.g.\ the mean squared gradient~\cite{Wei_2019}, do not register whether the gradient direction changes. In practice, the path length is set by experience~\cite{Liu_2019}, gravity matching areas are graded by heuristic feature parameters whose suitability depends on the track direction~\cite{Zhao_2024}, and informative paths are found by numerical planning~\cite{Reynaud_2010,Penumarti_2026,Kang_2026}.

This paper addresses the third challenge for scalar measurements of any mapped field. We model the path as known up to a constant horizontal position error, and prove a short-path law that bounds the Fisher information for the least-informed position component. That is, over paths of length $\leq L$ from a given start point, the largest weakest-direction path information satisfies $N_L=\kappa L^3+O(L^4)$ as $L\to0$, where $\kappa\propto(\|\nabla B\|\,\|\nabla\theta\|)^2$ at the start point, $B$ is the map, and $\theta$ is the gradient angle. Through the CRLB, $\kappa$ gives a local distance scale $\ell_{\rm req}\propto\kappa^{-1/3}$ at which the bound on the least-informed position component reaches a target $\sigma_{\rm req}$. The coefficient $\kappa$ requires only the map, and can be evaluated before a mission. It can be used to (i) screen map regions for localisation without a path search, (ii) choose the initial heading along $\nabla\theta$, and (iii) relate the sensor noise to the distance needed to reach $\sigma_{\rm req}$. We test the law on two aeromagnetic anomaly maps as worked examples, on which $\kappa$ ranks locations by the shortest straight-path length more closely than gradient magnitude does.

The results in this paper build on previous works in at least three ways: (i) the minimum eigenvalue of the summed gradient outer product, which Shi and Tomasi used to accept image windows for tracking~\cite{Shi_1994}, is the weakest-direction path information; (ii) the product of the map Hessian with the direction of motion, which appears in the observability matrix that Ramos \emph{et al.} used for guidance~\cite{Ramos_2022}, sets the leading coefficient; and (iii) the CRLB, which has served as a criterion for map suitability and for configuring the matching window~\cite{Reynaud_2010,Sang_2025}, converts $\kappa$ into a distance. We emphasise that the result is local and asymptotic. It sets a lower bound on the variance of any locally unbiased estimator, and the optimal heading holds to leading order in $L$.

The remainder of this paper is structured as follows. In Section~\ref{sec:information} we apply (i) to the path information and state the model idealisations. Section~\ref{sec:cubic} combines (i) with (ii) to derive the cubic law. We use (iii) in Section~\ref{sec:application} to convert $\kappa$ into $\ell_{\rm req}$ and test both on two aeromagnetic anomaly maps. The geometry of $\kappa$, an unknown measurement bias, and the map scale are canvassed briefly in Section~\ref{sec:discussion}, and Section~\ref{sec:conclusion} concludes.

\section{PATH INFORMATION}
\label{sec:information}

Let $B(\bm x)$ denote a continuously differentiable scalar map of horizontal position
$\bm x\in\mathbb R^2$. Let $\tilde{\bm p}(s)$ denote the inertial horizontal
position at arc length $0\le s\le L$, and let the true position be
${\bm p}_{\rm true}(s) = \tilde{\bm p}(s)+\delta\bm p$, where the horizontal position error
$\delta\bm p\in\mathbb R^2$ is constant over the interval, i.e.\ the
path is displaced rigidly (Fig.~\ref{fig1}). A vehicle samples $B$ with a scalar sensor at constant speed $v$, sampling rate $f$, and sample spacing $\Delta s = v/f$. The $j$-th measurement is
\begin{equation}
    z_j = B[\bm p_{\rm true}(s_j)] + n_j,
    \label{eq:measurement}
\end{equation}
where the errors $n_j \sim {\cal N}(0,\sigma^2)$ are independent, $s_j = j\Delta s$ for $0 \le j < M$, and $L=(M-1)\Delta s$.

Let $\bm{g}(s) = \nabla B[\bm p_{\rm true}(s)]$ denote the gradient of $B$ at $\bm p_{\rm true}(s)$. We define the path information as $\bm F = \int_0^L \bm{g}(s) \bm{g}(s)^\top\,{\rm d}s$ (units: $[B]^2$\,m$^{-1}$). The accumulated Fisher information $\bm{{\cal I}}_{M}$ (units: m$^{-2}$) for $\delta\bm p$ from $M$ samples is
\begin{equation}
    \bm{{\cal I}}_M
    = \frac{1}{\sigma^2}\sum_{j=0}^{M-1}\bm{g}_{j}\bm{g}_{j}^\top
    \approx \frac{\bm F}{\Delta s\,\sigma^2},
    \label{eq:pathinfo}
\end{equation}
for $\Delta s \ll L$, where $\bm{g}_{j} = \bm{g}(s_j)$. For $\bm{g}_{j}\neq \bm{0}$, each summand $\bm g_{j}\bm g_{j}^\top$ has rank one, and its null vector is the unit contour tangent $\hat{\bm g}_{\perp,j}$ at $\bm p_{\rm true}(s_j)$. The matrix $\bm{{\cal I}}_M$ has rank two if and only if the sampled gradients are not all collinear, i.e.\ $\det[\bm g_i\;\bm g_k] \ne 0$ for some $0 \le i < k < M$.\footnote{\label{fn:sampling}For $M$ equally spaced samples on a straight path, $\Delta s\,\lambda_{\min}(\sum_j\bm g_j\bm g_j^\top)$ exceeds $\lambda_{\min}(\bm F)$ by the factor $M(M+1)/(M-1)^2$ to leading order in $L$, e.g.\ 1.47 at $M=8$. The sampled information is then the product of a location-dependent factor and $(M-1)M(M+1)$, which increases with $M$, so the factor preserves the order of locations by required distance.}

Write the gradient as $\bm g(s)=\|\bm g(s)\|\,[\cos\theta(s),\sin\theta(s)]^\top$, where $\theta(s)$ is the gradient angle. The eigenvalues of $\bm F$ are then
\begin{equation}
    \lambda_{\max,\min}(\bm F) = \frac12\left(W \pm \sqrt{C^2+S^2}\right),
    \label{eq:spectrum}
\end{equation}
where $(W,C,S)=\int_0^L\|\bm g(s)\|^2\,(1,\cos2\theta,\sin2\theta)\,{\rm d}s$. The triangle inequality gives $|C+{\rm i}S|=|\int_0^L\|\bm g\|^2e^{2{\rm i}\theta}\,{\rm d}s|\le W$, with equality if and only if $\theta(s)$ is constant modulo $\pi$ wherever $\bm g(s)\neq\bm 0$. By Equation~(\ref{eq:spectrum}), $\lambda_{\min}(\bm F)=0$ if and only if the gradients along the path are collinear. We refer to $\lambda_{\min}(\bm F)$ as the weakest-direction path information. Gradient magnitude enters $W$, but it cannot make $\lambda_{\min}(\bm F)$ positive while the gradients remain collinear.

Equation~(\ref{eq:measurement}) is highly idealised in many respects. For example, it omits: (i) a position error that grows along the path because of a velocity or heading error, e.g.\ a random walk driven by white velocity noise $\bm\xi(t)$, viz.\ ${\rm d}(\delta\bm p)/{\rm d}t=\bm\xi(t)$; (ii) a constant or slowly varying measurement offset, e.g.\ a barometric altitude error in terrain-referenced navigation or diurnal variation in MagNav, which adds a bias to every $z_j$; and (iii) prior information on $\delta\bm p$ carried by a navigation filter. Points (i) and (ii) add a nuisance parameter $\bm\eta$. Let $\bm{{\cal I}}_{p\eta}$ and $\bm{{\cal I}}_{\eta\eta}$ denote the cross and nuisance blocks of the Fisher information for $(\delta\bm p,\bm\eta)$. Its block for $\delta\bm p$ is $\bm{{\cal I}}_M$, because $\partial z_j/\partial\delta\bm p=\bm g_j$ for any $\bm\eta$. The Fisher information for $\delta\bm p$ with $\bm\eta$ unknown is the Schur complement
\begin{equation}
    \bm{{\cal I}}_M^{(\eta)}
    = \bm{{\cal I}}_M-\bm{{\cal I}}_{p\eta}\,\bm{{\cal I}}_{\eta\eta}^{-1}\,\bm{{\cal I}}_{p\eta}^\top
    \preceq \bm{{\cal I}}_M,
    \label{eq:schur}
\end{equation}
with a pseudo-inverse if $\bm{{\cal I}}_{\eta\eta}$ is singular. It follows that $\lambda_{\min}(\bm{{\cal I}}_M^{(\eta)})\le\lambda_{\min}(\bm{{\cal I}}_M)$, and distances computed from Equation~(\ref{eq:pathinfo}) remain necessary distances under (i) and (ii). A prior on $\bm\eta$ reduces the loss in Equation~(\ref{eq:schur}) but cannot raise the information above $\bm{{\cal I}}_M$. Point (iii) acts in the opposite direction. In the Bayesian (Van Trees) bound on the mean-square error, prior information on $\delta\bm p$ adds to the expected value of $\bm{{\cal I}}_M$, and distances computed from Equation~(\ref{eq:pathinfo}) are then no longer necessary. We adopt the constant horizontal position error model because it is the simplest model whose Fisher information contains no parameters of the inertial error model, which is the property a map-suitability score requires. We return to point (ii) in Section~\ref{sec:discussion}.

\section{CUBIC SHORT-PATH LAW}
\label{sec:cubic}

Let $\bm x_0=\bm p_{\rm true}(0)$ denote the start of the path, and let $\bm g_0=\nabla B(\bm x_0)$ and $\bm H_0=\nabla^2B(\bm x_0)$ denote the gradient and the Hessian of the map at $\bm x_0$, respectively. For $\bm g_0\neq\bm 0$, the unit contour tangent at $\bm x_0$ is $\hat{\bm g}_{\perp,0}=(-g_{0,y},g_{0,x})^\top/\|\bm g_0\|$. At $\bm x_0$, $\nabla\theta$ is given by
\begin{equation}
    \nabla\theta(\bm x_0)
    =\frac{\bm H_0\hat{\bm g}_{\perp,0}}{\|\bm g_0\|},
    \label{eq:anglegrad}
\end{equation}
and the cubic coefficient $\kappa$ by
\begin{equation}
    \kappa(\bm x_0)
    =\frac{\|\bm H_0\hat{\bm g}_{\perp,0}\|^2}{12}
    =\frac{\|\bm g_0\|^2\|\nabla\theta(\bm x_0)\|^2}{12}.
    \label{eq:kappa}
\end{equation}
Equation~(\ref{eq:anglegrad}) follows from $\tan\theta=g_y/g_x$, which yields $\nabla\theta=(g_x\nabla g_y-g_y\nabla g_x)/\|\bm g\|^2$, and from the symmetry of $\bm H_0$.

Let ${\cal A}_L(\bm x_0)$ denote the set of paths $\gamma:[0,\ell]\to\mathbb R^2$, parameterised by arc length, with $\gamma(0)=\bm x_0$ and $\ell\le L$. The initial heading $\gamma'(0)$ and the curvature $\gamma''(s)$ are unconstrained. For each path $\gamma$, let $\bm F_\gamma$ denote $\bm F$ accumulated along $\gamma$. We define the largest weakest-direction path information as
\begin{equation}
    N_L(\bm x_0)
    =\sup_{\gamma\in{\cal A}_L(\bm x_0)}
      \lambda_{\min}(\bm F_\gamma).
    \label{eq:supremum}
\end{equation}
Extending a path adds a positive-semidefinite matrix $\bm P$ to $\bm F_\gamma$, and $\bm e^\top(\bm F_\gamma+\bm P)\bm e\ge\bm e^\top\bm F_\gamma\bm e$ for every unit vector $\bm e$. It follows that extending a path does not decrease $\lambda_{\min}(\bm F_\gamma)$, and the supremum in Equation~(\ref{eq:supremum}) may be taken over paths of length exactly $L$.

\begin{theorem}\label{thm:cubic}
Let $B$ have continuous derivatives through third order near $\bm x_0$, with $\bm g_0\neq\bm 0$ and $\rho_0=\|\nabla\theta(\bm x_0)\|\neq0$. Then, as $L\rightarrow0$:
\begin{enumerate}
\item[(i)] the largest weakest-direction path information satisfies
\begin{equation}
    N_L(\bm x_0)=\kappa(\bm x_0)L^3+O(L^4);
    \label{eq:cubic}
\end{equation}
\item[(ii)] the straight path of length $L$ with heading
\begin{equation}
    \bm u_0
    =\pm\frac{\nabla\theta(\bm x_0)}{\rho_0}
    =\pm\frac{\bm H_0\hat{\bm g}_{\perp,0}}
              {\|\bm H_0\hat{\bm g}_{\perp,0}\|}
    \label{eq:optimalheading}
\end{equation}
attains $\lambda_{\min}(\bm F_\gamma)=\kappa L^3+O(L^4)$;
\item[(iii)] the straight path of length $L$ with unit heading $\bm u$ at angle $\phi$ to $\bm u_0$ has
\begin{equation}
    \begin{aligned}
    \lambda_{\min}(\bm F_\gamma)
    &=\frac{(\hat{\bm g}_{\perp,0}^\top \bm H_0\bm u)^2}{12}L^3+O(L^4)
    \\
    &=\kappa\cos^2\!\phi\,L^3+O(L^4).
    \end{aligned}
    \label{eq:heading}
\end{equation}
\end{enumerate}
\end{theorem}

\begin{IEEEproof}
Let $\gamma\in{\cal A}_L(\bm x_0)$ have length $L$. Let $\vartheta(s)=\theta[\gamma(s)]$ denote the gradient angle along $\gamma$ (chosen to be continuous in $s$), and let ${\rm Var}_s$ denote the variance under the uniform measure on $[0,L]$. Every path in ${\cal A}_L(\bm x_0)$ lies within distance $L$ of $\bm x_0$, where, for small $L$, $\bm g\ne\bm 0$ and the third derivatives of $B$ are bounded, so each $O(L^k)$ term below is bounded by $cL^k$ with $c$ independent of $\gamma$.

First, we choose the origin of the angle $\theta$ to give $S=0$ and $C\ge0$ in Equation~(\ref{eq:spectrum}). This choice leaves $W$ and $C^2+S^2$ unchanged. The weakest-direction path information then takes the exact form
\begin{equation}
    \lambda_{\min}(\bm F_\gamma)
    =\tfrac12(W-C)
    =\int_0^L\|\bm g(s)\|^2\sin^2\vartheta(s)\,{\rm d}s.
    \label{eq:exactmin}
\end{equation}

Second, we expand Equation~(\ref{eq:exactmin}) in $L$. Because $\gamma$ is parameterised by arc length, $|{\rm d}\vartheta/{\rm d}s|=|\nabla\theta^\top\gamma'(s)|\le\rho_0+O(L)$. The values of $2\vartheta$ on $[0,L]$ lie in an arc of width $O(L)$, which is less than $\pi$ for small $L$. The argument of $C+{\rm i}S=\int_0^L\|\bm g\|^2e^{2{\rm i}\vartheta}\,{\rm d}s$ lies in the same arc, and $S=0$ and $C\ge0$ set that argument to zero. This gives $\vartheta=O(L)$ on $[0,L]$. Expanding $S=0$ with $\|\bm g\|^2=\|\bm g_0\|^2+O(L)$ gives $\int_0^L\vartheta\,{\rm d}s=O(L^3)$. The identity $\int_0^L\vartheta^2\,{\rm d}s=L\,{\rm Var}_s(\vartheta)+L^{-1}(\int_0^L\vartheta\,{\rm d}s)^2$ then gives $\int_0^L\vartheta^2\,{\rm d}s=L\,{\rm Var}_s(\vartheta)+O(L^5)$. Substituting $\sin^2\vartheta=\vartheta^2+O(L^4)$ and $\|\bm g\|^2=\|\bm g_0\|^2+O(L)$ into Equation~(\ref{eq:exactmin}) gives
\begin{equation}
    \lambda_{\min}(\bm F_\gamma)
    =\|\bm g_0\|^2L\,{\rm Var}_s(\vartheta)+O(L^4).
    \label{eq:variance}
\end{equation}

Third, we bound the variance. The bound on $|{\rm d}\vartheta/{\rm d}s|$ gives $|\vartheta(s)-\vartheta(t)|\le[\rho_0+O(L)]\,|s-t|$, and $\int_0^L\!\int_0^L(s-t)^2\,{\rm d}s\,{\rm d}t=L^4/6$. Together these give
\begin{equation}
    \begin{aligned}
    {\rm Var}_s(\vartheta)
    &=\frac{1}{2L^2}\int_0^L\!\!\int_0^L
      [\vartheta(s)-\vartheta(t)]^2\,{\rm d}s\,{\rm d}t
    \\
    &\le\frac{\rho_0^2L^2}{12}+O(L^3).
    \end{aligned}
    \label{eq:variancebound}
\end{equation}
Equations~(\ref{eq:kappa}), (\ref{eq:variance}), and (\ref{eq:variancebound}) give $\lambda_{\min}(\bm F_\gamma)\le\kappa L^3+O(L^4)$ for every $\gamma\in{\cal A}_L(\bm x_0)$, and $N_L\le\kappa L^3+O(L^4)$. For the straight path with unit heading $\bm u$, $\vartheta(s)=\vartheta(0)+[\nabla\theta(\bm x_0)^\top\bm u]\,s+O(s^2)$ and ${\rm Var}_s(\vartheta)=[\nabla\theta(\bm x_0)^\top\bm u]^2L^2/12+O(L^3)$. Equations~(\ref{eq:anglegrad}), (\ref{eq:kappa}), and (\ref{eq:variance}) then give Equation~(\ref{eq:heading}), which proves (iii). Setting $\bm u=\bm u_0$ gives $\kappa L^3+O(L^4)$, which proves (ii) and, with the upper bound, (i).
\end{IEEEproof}

The bound in Equation~(\ref{eq:variancebound}) depends only on arc length and $\rho_0$, and holds for every path in ${\cal A}_L(\bm x_0)$. Curvature cannot raise the cubic coefficient above $\kappa$. Let $o(L^k)$ denote a term whose ratio to $L^k$ tends to zero as $L\to0$. A path with $\sup_s|\vartheta(s)-\vartheta(0)\mp\rho_0s|=o(L)$, e.g.\ a path whose tangent stays close to $\pm\bm u_0$, has $\lambda_{\min}(\bm F_\gamma)=\kappa L^3+o(L^3)$. The strongest-direction path information is $\lambda_{\max}(\bm F_\gamma)=W-\lambda_{\min}(\bm F_\gamma)=\|\bm g_0\|^2L+O(L^2)$, and Equation~(\ref{eq:variance}) then gives
\begin{equation}
    \frac{\lambda_{\min}(\bm F_\gamma)}{\lambda_{\max}(\bm F_\gamma)}
    ={\rm Var}_s(\vartheta)+O(L^3).
    \label{eq:ratio}
\end{equation}
In Equation~(\ref{eq:cubic}), one power of $L$ comes from the accumulated squared gradient $\|\bm g_0\|^2L$, and two come from the variance $\rho_0^2L^2/12$ of the gradient angle along $\pm\bm u_0$. If $\bm H_0\hat{\bm g}_{\perp,0}=\bm0$, i.e.\ $\rho_0=0$, the cubic coefficient vanishes and higher map derivatives determine the leading order.

\section{TEST ON TWO AEROMAGNETIC ANOMALY MAPS}
\label{sec:application}

In this section, we test Theorem~\ref{thm:cubic} and the distance scale of Equation~(\ref{eq:pathlength}) on two aeromagnetic anomaly maps. In Section~\ref{sec:maps}, we describe the maps, the smoothing, and the lattice of locations. The test has four parts. First, in Section~\ref{sec:law}, we compare the weakest-direction path information along $+\bm u_0$ with its leading term $\kappa L^3$ at every retained location. Second, in Section~\ref{sec:distance}, we compare the distance scale with the shortest straight path, found by a search over headings, at which the sampled Cram\'er--Rao bound reaches a target standard deviation, and compare the heading $\pm\bm u_0$ with the best of the searched headings. Third, in Section~\ref{sec:ranking}, we rank the retained locations by $\kappa$ and by the gradient magnitude, and compare both rankings with that shortest path, with one analysis choice changed at a time. Fourth, in Section~\ref{sec:estimator}, we simulate a maximum-likelihood estimator of $\delta\bm p$, and compare the distance it requires with that shortest path.

\subsection{Maps and Lattice}
\label{sec:maps}

\begin{figure}[t]
    \centering
    \includegraphics[width=\columnwidth]{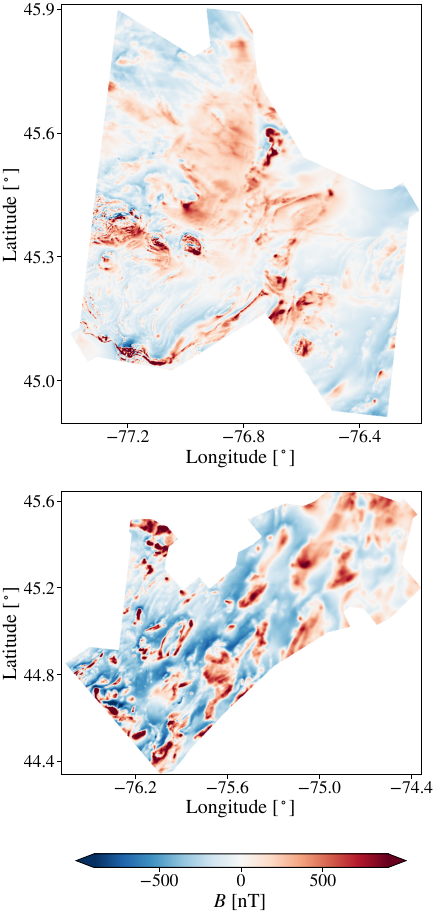}
    \caption{The \texttt{Renfrew\_395} (top panel) and \texttt{Eastern\_395} (bottom panel) aeromagnetic anomaly maps $B$ (units: nT)~\cite{Gnadt_2020}, before smoothing. Unsurveyed cells are white. Geodetic longitude (units: degrees) is reported on the horizontal axis and geodetic latitude (units: degrees) on the vertical axis. The colour bar spans $\pm900$~nT, the 99th percentile of $|B|$ over both maps, and larger values are saturated. The retained locations of Fig.~\ref{fig3} lie at least 3.5~km inside the surveyed area (Section~\ref{sec:maps}).}
    \label{fig2}
\end{figure}

The \texttt{Renfrew\_395} and \texttt{Eastern\_395} aeromagnetic anomaly maps are distributed with the MagNav.jl Julia package~\cite{Gnadt_2020}.\footnote{\url{https://github.com/MIT-AI-Accelerator/MagNav.jl}} The data were collected near Ottawa, Canada, by Sander Geophysics Ltd.,\footnote{\url{https://sgl.com}} and upward continued to a level surface 395~m above the WGS-84 ellipsoid. The surveyed areas of \texttt{Renfrew\_395} and \texttt{Eastern\_395} cover 6150 and 12,397~km$^2$, on map cells measuring $29.2\times29.5$~m and $29.6\times29.8$~m, respectively. Fig.~\ref{fig2} shows both maps before smoothing.

We smooth both maps with a Gaussian kernel of standard deviation $w=300$~m, and vary $w$ in Table~\ref{tab:robust}. We compute the gradient $\nabla B$ and the Hessian $\bm H$ of the smoothed map in the Fourier domain. We evaluate every result on a lattice of every 16th map cell, i.e.\ $\sim470$~m apart. We retain the 20,894 and 43,283 lattice locations on \texttt{Renfrew\_395} and \texttt{Eastern\_395}, respectively, whose paths of up to 2~km stay at least $5w=1.5$~km inside the surveyed area, i.e.\ locations at least 3.5~km from its edge.

\subsection{Cubic Law}
\label{sec:law}

We evaluate the ratio $\lambda_{\min}(\bm F_\gamma)/(\kappa L^3)$ at every retained location along the heading $+\bm u_0$ of Equation~(\ref{eq:optimalheading}), at $L=30$, 75, and 300~m. At $L=30$~m, the median ratio equals 0.996 and 0.998 on \texttt{Renfrew\_395} and \texttt{Eastern\_395}, respectively. At $L=75$~m, it equals 0.975 and 0.990. At $L=300$~m, it equals 0.715 and 0.873. The spread of the ratio grows with $L$, consistent with the $O(L^4)$ remainder in Equation~(\ref{eq:cubic}). On \texttt{Renfrew\_395}, the 10th and 90th percentiles at $L=300$~m equal 0.139 and 2.607. That is, the leading term predicts the median to within 3\% up to $L=75$~m, and does not predict individual locations at $L=300$~m.

\subsection{Required Distance}
\label{sec:distance}

The Cram\'er--Rao bound states that $\bm{{\cal I}}_M^{-1}$ is a lower bound on the covariance of any locally unbiased estimator of $\delta\bm p$ [see Chapter~3 of \cite{Kay_1993}]. Let $\sigma_{\rm worst}^2$ denote the largest horizontal directional variance of the estimator. We refer to the eigenvector of $\bm{{\cal I}}_M$ with eigenvalue $\lambda_{\min}(\bm{{\cal I}}_M)$ as the worst direction, because the bound on the directional variance is largest along it. By Equation~(\ref{eq:pathinfo}),
\begin{equation}
    \sigma_{\rm worst}^2
    \ge\frac{1}{\lambda_{\min}(\bm{{\cal I}}_M)}
    \approx\frac{\Delta s\,\sigma^2}{\lambda_{\min}(\bm F_\gamma)}.
    \label{eq:worstcrlb}
\end{equation}
Let $\sigma_{\rm req}$ denote a target standard deviation. We define the local distance scale $\ell_{\rm req}(\bm x)$ as the path length at which the bound in Equation~(\ref{eq:worstcrlb}), with $\lambda_{\min}(\bm F_\gamma)=\kappa L^3$ from Equation~(\ref{eq:cubic}), equals $\sigma_{\rm req}^2$, viz.
\begin{equation}
    \ell_{\rm req}(\bm x)
    =\left[\frac{\Delta s\,\sigma^2}
    {\sigma_{\rm req}^2\kappa(\bm x)}\right]^{1/3}.
    \label{eq:pathlength}
\end{equation}
Equation~(\ref{eq:pathlength}) approximates the required distance to leading order in $\ell_{\rm req}$, i.e.\ for a short path over which the local map geometry changes little and $\Delta s\ll\ell_{\rm req}$.

\begin{figure*}[t]
    \centering
    \includegraphics[width=\textwidth]{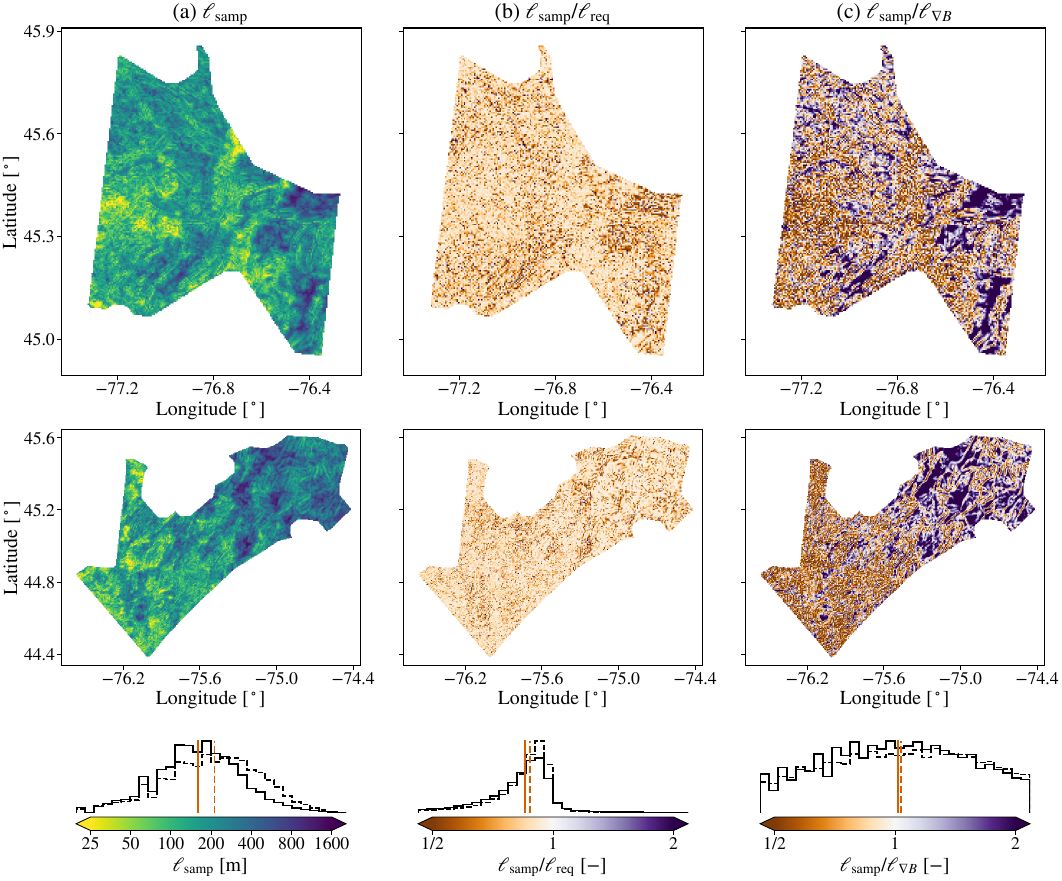}
    \caption{Sampled distance $\ell_{\rm samp}$ and its ratios to $\ell_{\rm req}$ and $\ell_{\nabla B}$ at the retained locations of the \texttt{Renfrew\_395} (top row) and \texttt{Eastern\_395} (middle row) aeromagnetic anomaly maps, for $\sigma_{\rm req}=100$~m and $w=300$~m. We plot $\ell_{\rm samp}$ (units: m), $\ell_{\rm samp}/\ell_{\rm req}$ (dimensionless), and $\ell_{\rm samp}/\ell_{\nabla B}$ (dimensionless) in columns (a), (b), and (c), respectively, where $\ell_{\nabla B}$ is the distance predicted from $\|\nabla B\|$. In column (a), yellow and purple denote short and long $\ell_{\rm samp}$, respectively. In columns (b) and (c), white denotes a ratio of unity, i.e.\ a predicted distance equal to $\ell_{\rm samp}$, and orange and purple denote ratios below and above unity, respectively. The colour scales are logarithmic. Column (b) is pale over both maps, i.e.\ $\ell_{\rm req}$, computed from $\kappa$ without a path search or a fitted parameter, predicts $\ell_{\rm samp}$ with no spatial pattern. Column (c) departs from white in patches, i.e.\ $\ell_{\nabla B}$, although fitted to $\ell_{\rm samp}$, misses where the gradient axis turns slowly (Section~\ref{sec:distance}). The distributions of the three quantities are plotted in the bottom row, for \texttt{Renfrew\_395} (solid curves) and \texttt{Eastern\_395} (dashed curves). Overplotted as vertical, orange lines are the medians of each distribution. Geodetic longitude (units: degrees) is reported on the horizontal axis and geodetic latitude (units: degrees) on the vertical axis.}
    \label{fig3}
\end{figure*}

Let $\bm{{\cal I}}_M(\bm x,\bm u)$ denote the Fisher information in Equation~(\ref{eq:pathinfo}) accumulated along the straight path $\bm x+s\bm u$, $0\le s\le(M-1)\Delta s$, from location $\bm x$ with unit heading $\bm u$. We compare Equation~(\ref{eq:pathlength}) with the sampled distance
\begin{equation}
    \begin{aligned}
    \ell_{\rm samp}(\bm x)=\min_{\bm u\in{\cal U}}\,\min\bigl\{&(M-1)\Delta s:\\
    &\lambda_{\min}(\bm{{\cal I}}_M(\bm x,\bm u))\ge\sigma_{\rm req}^{-2}\bigr\},
    \end{aligned}
    \label{eq:lsamp}
\end{equation}
where ${\cal U}$ is the set of 72 unit headings at $5^\circ$ intervals. That is, $\ell_{\rm samp}$ is the shortest straight path, over the headings in ${\cal U}$ and the number of samples $M$, at which the sampled bound in Equation~(\ref{eq:worstcrlb}) reaches $\sigma_{\rm req}$. By Equation~(\ref{eq:lsamp}), $\ell_{\rm samp}$ is a multiple of $\Delta s$. Each path extends up to 2~km. We take $v=68\,{\rm m\,s^{-1}}$, $f=10$~Hz and $\sigma=3$~nT, i.e.\ $\Delta s=6.8$~m, and $\sigma_{\rm req}=100$~m. By Equations~(\ref{eq:pathinfo}), (\ref{eq:pathlength}), and (\ref{eq:lsamp}), $\sigma$ and $\sigma_{\rm req}$ enter $\ell_{\rm req}$ and $\ell_{\rm samp}$ through the ratio $\sigma/\sigma_{\rm req}$, and $v$ and $f$ through $\Delta s$. That is, the rows of Table~\ref{tab:robust} with $\sigma_{\rm req}=50$ and 200~m also give the results for $\sigma=6$ and 1.5~nT at $\sigma_{\rm req}=100$~m. Every location reaches $\sigma_{\rm req}$ within 2~km. The median $\ell_{\rm samp}$ equals 163.2 and 210.8~m on \texttt{Renfrew\_395} and \texttt{Eastern\_395}, respectively, i.e.\ 2.4 and 3.1~s of flight.

Let $\ell_{\nabla B}(\bm x)=h(\|\nabla B(\bm x)\|)$ denote the distance predicted from the gradient magnitude, where $h$ is the non-increasing function that minimises $\sum_i[\log\ell_{\rm samp}(\bm x_i)-\log h(\|\nabla B(\bm x_i)\|)]^2$ over the retained locations $\bm x_i$ of each map, i.e.\ an isotonic regression. That is, $\ell_{\nabla B}$ converts $\|\nabla B\|$ into a distance, calibrated against $\ell_{\rm samp}$.

The three distances $\ell_{\rm req}$, $\ell_{\rm samp}$, and $\ell_{\nabla B}$ estimate how far a vehicle must travel from $\bm x$ before the worst-direction standard deviation reaches $\sigma_{\rm req}$. That is, a small distance marks a location at which a short path localises the vehicle to $\sigma_{\rm req}$. A navigator seeks locations at which $\ell_{\rm samp}$ is as small as possible. Specifically, $\ell_{\rm samp}$ follows from a search over headings and path lengths by Equation~(\ref{eq:lsamp}), whereas $\ell_{\rm req}$ follows from $\kappa$ by Equation~(\ref{eq:pathlength}), and $\ell_{\nabla B}$ from $\|\nabla B\|$ by the isotonic fit above. The two predictions $\ell_{\rm req}$ and $\ell_{\nabla B}$ are judged by how closely they match $\ell_{\rm samp}$, i.e.\ by how close the ratios $\ell_{\rm samp}/\ell_{\rm req}$ and $\ell_{\rm samp}/\ell_{\nabla B}$ lie to unity. Ultimately, a large $\kappa$, i.e.\ a small $\ell_{\rm req}$, identifies the locations with small $\ell_{\rm samp}$ without the search of Equation~(\ref{eq:lsamp}).

In Fig.~\ref{fig3}, we plot $\ell_{\rm samp}$ (units: m), $\ell_{\rm samp}/\ell_{\rm req}$ (dimensionless), and $\ell_{\rm samp}/\ell_{\nabla B}$ (dimensionless) in columns (a), (b), and (c), respectively, at the retained locations of \texttt{Renfrew\_395} (top row) and \texttt{Eastern\_395} (middle row). Fig.~\ref{fig3} has three key features. First, $\ell_{\rm samp}$ changes on scales of a few kilometres. Its 5th and 95th percentiles are 54.4 and 476.0~m on \texttt{Renfrew\_395}, and 54.4 and 632.4~m on \texttt{Eastern\_395}. Second, $\ell_{\rm samp}/\ell_{\rm req}$ shows no spatial pattern. Its correlation between neighbouring lattice locations equals 0.02 and 0.12 on \texttt{Renfrew\_395} and \texttt{Eastern\_395}, respectively. Its median equals 0.865 and 0.890. It lies within a factor 1.5 of unity at 85.0\% and 89.6\% of locations. That is, $\ell_{\rm req}$ predicts $\ell_{\rm samp}$ equally well over the whole of both surveys. Third, $\ell_{\rm samp}/\ell_{\nabla B}$ varies in patches. Its neighbour correlation equals 0.45 and 0.65. It lies within a factor 1.5 of unity at 55.3\% and 50.0\% of locations. In the patches with $\ell_{\rm samp}/\ell_{\nabla B}>2$, the median $\|\nabla\theta\|$ equals $0.28\times10^{-3}$ and $0.16\times10^{-3}$~rad\,m$^{-1}$, against $1.50\times10^{-3}$ and $0.87\times10^{-3}$~rad\,m$^{-1}$ elsewhere, indicating that $\|\nabla B\|$ underestimates $\ell_{\rm samp}$ where the gradient axis turns slowly.

The ratio $\ell_{\rm samp}/\ell_{\rm req}$ is below unity at 93.0\% and 94.5\% of locations on \texttt{Renfrew\_395} and \texttt{Eastern\_395}, respectively [see the bottom row of Fig.~\ref{fig3}(b)], and decreases as $\ell_{\rm req}$ increases. On \texttt{Renfrew\_395}, its median equals 0.91 for $\ell_{\rm req}<75$~m and 0.37 for $\ell_{\rm req}>1.2$~km. On \texttt{Eastern\_395}, the corresponding medians are 0.91 and 0.46. By Equation~(\ref{eq:worstcrlb}), $\ell_{\rm samp}$ is the necessary distance along these paths, and $\ell_{\rm req}$ is its leading-order scale. Two effects shorten $\ell_{\rm samp}$ relative to $\ell_{\rm req}$, viz.\ (i) the sampling factor of footnote~\ref{fn:sampling} on short paths, and (ii) the higher-order terms of Equation~(\ref{eq:cubic}) on long paths. Dividing the ratio by the sampling factor gives a median of 1.03 for $\ell_{\rm req}<75$~m on both maps.

Let $\ell_{\bm u_0}$ denote the shorter of the two distances, along $+\bm u_0$ and $-\bm u_0$ of Equation~(\ref{eq:optimalheading}), at which the sampled bound in Equation~(\ref{eq:worstcrlb}) reaches $\sigma_{\rm req}$. The heading $\pm\bm u_0$ attains $\ell_{\rm samp}$, i.e.\ $\ell_{\bm u_0}=\ell_{\rm samp}$, at 55.6\% and 59.6\% of locations on \texttt{Renfrew\_395} and \texttt{Eastern\_395}, respectively. It gives $\ell_{\bm u_0}\le1.1\,\ell_{\rm samp}$ at 81.7\% and 87.5\% of locations, and $\ell_{\bm u_0}\le1.5\,\ell_{\rm samp}$ at 95.6\% and 97.2\%. At 14 and 54 locations, the path along $\pm\bm u_0$ does not reach $\sigma_{\rm req}$ within 2~km. Along a fixed axis, viz.\ $0^\circ$, $45^\circ$, $90^\circ$, or $135^\circ$ clockwise from north in the shorter of its two directions, the median distance at which the bound reaches $\sigma_{\rm req}$ lies between 217.6 and 238.0~m on \texttt{Renfrew\_395}, against 170.0~m along $\pm\bm u_0$, i.e.\ 28\% to 40\% longer. On \texttt{Eastern\_395}, it is 26\% to 48\% longer. That is, the heading $\pm\bm u_0$ of Theorem~\ref{thm:cubic}(ii) is near-optimal among the 72 straight headings.

\subsection{Ranking of Locations}
\label{sec:ranking}

\begin{table}[t]
\caption{Median distance ratio $\ell_{\rm samp}/\ell_{\rm req}$ and Spearman correlations $\rho_{\rm S}$ on the \texttt{Renfrew\_395} (R) and \texttt{Eastern\_395} (E) aeromagnetic anomaly maps. The first row is the baseline ($w=300$~m, $\sigma_{\rm req}=100$~m, 395~m altitude), on 20,894 (R) and 43,283 (E) locations. The second row uses \texttt{Renfrew\_555}. The third and fourth rows vary $\sigma_{\rm req}$. The last three rows vary $w$, on the 18,146 (R) and 38,485 (E) locations retained at $w=600$~m. Dashes mark entries not computed.}
\label{tab:robust}
\centering
\footnotesize
\setlength{\tabcolsep}{3pt}
\begin{tabular}{@{}l*{6}{c}@{}}
\toprule
 & \multicolumn{2}{c}{$\ell_{\rm samp}/\ell_{\rm req}$} & \multicolumn{2}{c}{$\rho_{\rm S}(\kappa,\ell_{\rm samp})$} & \multicolumn{2}{c}{$\rho_{\rm S}(\|\nabla B\|,\ell_{\rm samp})$}\\
\cmidrule(lr){2-3}\cmidrule(lr){4-5}\cmidrule(l){6-7}
 & R & E & R & E & R & E\\
\midrule
Baseline & 0.865 & 0.890 & $-0.96$ & $-0.97$ & $-0.52$ & $-0.56$\\
\midrule
Altitude 555~m & 0.861 & -- & $-0.94$ & -- & $-0.50$ & --\\
\midrule
$\sigma_{\rm req}=50$~m & 0.840 & 0.866 & $-0.91$ & $-0.95$ & $-0.58$ & $-0.59$\\
$\sigma_{\rm req}=200$~m & 0.881 & 0.903 & $-0.98$ & $-0.98$ & $-0.49$ & $-0.55$\\
\midrule
$w=150$~m & 0.848 & 0.882 & $-0.96$ & $-0.97$ & $-0.58$ & $-0.58$\\
$w=300$~m & 0.864 & 0.891 & $-0.95$ & $-0.97$ & $-0.51$ & $-0.56$\\
$w=600$~m & 0.873 & 0.901 & $-0.94$ & $-0.97$ & $-0.47$ & $-0.50$\\
\bottomrule
\end{tabular}
\end{table}

Let $\rho_{\rm S}(x,y)$ denote the Spearman rank correlation coefficient between $x$ and $y$. A score that orders the retained locations exactly by $\ell_{\rm samp}$ gives $\rho_{\rm S}=-1$, because a large score predicts a short distance. The correlation $\rho_{\rm S}(\kappa,\ell_{\rm samp})$ equals $-0.96$ and $-0.97$ on \texttt{Renfrew\_395} and \texttt{Eastern\_395}, respectively. The correlation $\rho_{\rm S}(\|\nabla B\|,\ell_{\rm samp})$ equals $-0.52$ and $-0.56$. Of the tenth of locations with the largest $\kappa$, 92.6\% and 94.2\% lie in the tenth with the shortest $\ell_{\rm samp}$. Of the tenth of locations with the largest $\|\nabla B\|$, 44.4\% and 43.0\% lie in the tenth with the shortest $\ell_{\rm samp}$. By Equation~(\ref{eq:kappa}), $\kappa=(\|\nabla B\|\,\|\nabla\theta\|)^2/12$. The correlation $\rho_{\rm S}(\|\nabla\theta\|,\ell_{\rm samp})$ equals $-0.54$ and $-0.56$, i.e.\ it is similar to that of $\|\nabla B\|$. For $\ell_{\rm req}\ge600$~m, $\rho_{\rm S}(\kappa,\ell_{\rm samp})$ equals $-0.37$ and $-0.46$, against $-0.45$ and $-0.27$ for $\|\nabla B\|$, consistent with Theorem~\ref{thm:cubic} holding to leading order in $L$.

We test the sensitivity of these results to three analysis choices, viz.\ (i) the altitude, using the \texttt{Renfrew\_555} map, i.e.\ the same survey upward continued to 555~m; (ii) the target standard deviation, $\sigma_{\rm req}=50$ and 200~m; and (iii) the smoothing width, $w=150$ and 600~m. In Table~\ref{tab:robust}, we list the median $\ell_{\rm samp}/\ell_{\rm req}$ and the Spearman correlations with one choice changed at a time. The smoothing rows use the 18,146 and 38,485 locations retained at $w=600$~m, because the margin $5w$ grows with $w$. Table~\ref{tab:robust} repeats the $w=300$~m values on these locations. We draw the reader's attention to two features of Table~\ref{tab:robust}. First, $\rho_{\rm S}(\kappa,\ell_{\rm samp})$ lies between $-0.91$ and $-0.98$ in every row. Second, $\rho_{\rm S}(\|\nabla B\|,\ell_{\rm samp})$ lies between $-0.47$ and $-0.59$, and the median $\ell_{\rm samp}/\ell_{\rm req}$ between 0.840 and 0.903. That is, the ordering of locations by $\kappa$ and the ratio $\ell_{\rm samp}/\ell_{\rm req}$ are insensitive to the three choices. The distances scale with $w$. On \texttt{Renfrew\_395}, the median $\ell_{\rm samp}$ equals 108.8, 163.2, and 306.0~m at $w=150$, 300, and 600~m, respectively. On \texttt{Eastern\_395}, it equals 176.8, 210.8, and 285.6~m.

\subsection{Maximum-Likelihood Estimator}
\label{sec:estimator}

The Cram\'er--Rao bound in Equation~(\ref{eq:worstcrlb}) is local, i.e.\ it depends on the Fisher information at the true $\delta\bm p$ only. An estimator can select an offset far from the true $\delta\bm p$ at which the map predicts similar measurements, and the bound does not account for such an ambiguity. In this experiment, we test $\ell_{\rm samp}$ with the maximum-likelihood estimator, viz.
\begin{equation}
    \widehat{\delta\bm p}=\arg\min_{\delta\bm p\in{\cal D}}\sum_{j=0}^{M-1}
    \bigl\{z_j-B[\tilde{\bm p}(s_j)+\delta\bm p]\bigr\}^2,
    \label{eq:ml}
\end{equation}
where ${\cal D}$ is a square grid of half-width $a$ and spacing 10~m, centred on the true $\delta\bm p$. Equation~(\ref{eq:ml}) is a batch map-matching estimator with a squared-error cost. For the errors $n_j$ of Equation~(\ref{eq:measurement}), it gives the maximum a posteriori estimate for a uniform prior on ${\cal D}$. That is, $a$ bounds the prior position error, e.g.\ the error of the inertial position $\tilde{\bm p}$ at the start of the path. Centring ${\cal D}$ on the true $\delta\bm p$ favours the estimator, because the true $\delta\bm p$ is then a grid point of ${\cal D}$. We persevere with this idealisation, because it requires no parameters of the inertial error model, as in Section~\ref{sec:information}.

We simulate Equation~(\ref{eq:ml}) at 200 retained locations per map, drawn at random. At each location, the path follows the heading in ${\cal U}$ that attains $\ell_{\rm samp}$ in Equation~(\ref{eq:lsamp}). We generate $K=500$ measurement records by Equation~(\ref{eq:measurement}) with $\delta\bm p=\bm 0$ on the smoothed map, and evaluate Equation~(\ref{eq:ml}) on the same map, i.e.\ without map error. Let $\hat{\bm C}=K^{-1}\sum_{k=1}^{K}\widehat{\delta\bm p}_k\widehat{\delta\bm p}_k^\top$ denote the mean-square error matrix of the $K$ estimates $\widehat{\delta\bm p}_k$. We estimate $\sigma_{\rm worst}$ in Equation~(\ref{eq:worstcrlb}) by $\lambda_{\max}(\hat{\bm C})^{1/2}$, which includes any bias of the estimator. We define $\ell_{\rm ML}$ as the shortest length of the same straight path at which $\lambda_{\max}(\hat{\bm C})^{1/2}\le\sigma_{\rm req}$, evaluated from $\ell_{\rm samp}$ upwards at lengths 5\% apart. That is, $\ell_{\rm ML}$ is the distance that the estimator of Equation~(\ref{eq:ml}) requires, and $\ell_{\rm ML}/\ell_{\rm samp}$ measures how far the estimator falls short of the bound. We repeat the test for $a=300$, 400, 500, 700, and 1000~m, with the same measurement records at every $a$.

\begin{table}[t]
\caption{Distance $\ell_{\rm ML}$ at which a maximum-likelihood estimator
reaches $\sigma_{\rm req}$, and its rank correlation with $\kappa$ and
$\|\nabla B\|$. Median ratio
$\ell_{\rm ML}/\ell_{\rm samp}$ and Spearman correlations $\rho_{\rm S}$ on 200
random retained locations of the \texttt{Renfrew\_395} (R) and
\texttt{Eastern\_395} (E) aeromagnetic anomaly maps, for the half-width $a$ of the
search grid ${\cal D}$ of Equation~(\ref{eq:ml}). Each entry follows from $K=500$
simulated measurement records per location. The first row gives the correlations
with $\ell_{\rm samp}$ on the same locations, for comparison with
Table~\ref{tab:robust}.}
\label{tab:estimator}
\centering
\footnotesize
\setlength{\tabcolsep}{3pt}
\begin{tabular}{@{}l*{6}{c}@{}}
\toprule
 & \multicolumn{2}{c}{$\ell_{\rm ML}/\ell_{\rm samp}$}
 & \multicolumn{2}{c}{$\rho_{\rm S}(\kappa,\cdot)$}
 & \multicolumn{2}{c}{$\rho_{\rm S}(\|\nabla B\|,\cdot)$}\\
\cmidrule(lr){2-3}\cmidrule(lr){4-5}\cmidrule(l){6-7}
 & R & E & R & E & R & E\\
\midrule
$\ell_{\rm samp}$ & -- & -- & $-0.97$ & $-0.98$ & $-0.49$ & $-0.56$\\
\midrule
$\ell_{\rm ML}$, $a=300$~m  & 1.11 & 1.06 & $-0.85$ & $-0.97$ & $-0.52$ & $-0.53$\\
$\ell_{\rm ML}$, $a=400$~m  & 1.14 & 1.07 & $-0.76$ & $-0.92$ & $-0.55$ & $-0.52$\\
$\ell_{\rm ML}$, $a=500$~m  & 1.22 & 1.08 & $-0.67$ & $-0.83$ & $-0.55$ & $-0.51$\\
$\ell_{\rm ML}$, $a=700$~m  & 1.70 & 1.11 & $-0.57$ & $-0.76$ & $-0.56$ & $-0.50$\\
$\ell_{\rm ML}$, $a=1000$~m & 2.32 & 1.27 & $-0.49$ & $-0.60$ & $-0.48$ & $-0.46$\\
\bottomrule
\end{tabular}
\end{table}

In Table~\ref{tab:estimator}, we list the median $\ell_{\rm ML}/\ell_{\rm samp}$ and the Spearman correlations of $\kappa$ and $\|\nabla B\|$ with $\ell_{\rm ML}$ for each $a$. Table~\ref{tab:estimator} has three key features. First, for $a=300$~m, the estimator comes close to the bound. The median $\ell_{\rm ML}/\ell_{\rm samp}$ equals 1.11 and 1.06 on \texttt{Renfrew\_395} and \texttt{Eastern\_395}, respectively. Second, $\ell_{\rm ML}/\ell_{\rm samp}$ grows with $a$, to a median of 2.32 and 1.27 for $a=1000$~m. Third, $\rho_{\rm S}(\kappa,\ell_{\rm ML})$ weakens as $a$ grows, from $-0.85$ and $-0.97$ for $a=300$~m to $-0.49$ and $-0.60$ for $a=1000$~m. By contrast, $\rho_{\rm S}(\|\nabla B\|,\ell_{\rm ML})$ lies between $-0.46$ and $-0.56$ for every $a$. The difference $\rho_{\rm S}(\kappa,\ell_{\rm ML})-\rho_{\rm S}(\|\nabla B\|,\ell_{\rm ML})$ is negative, with a 90\% bootstrap interval that excludes zero, for $a\le500$~m on \texttt{Renfrew\_395} and for every $a$ on \texttt{Eastern\_395}. On \texttt{Renfrew\_395}, the interval includes zero for $a=700$ and 1000~m.

One possible explanation for the second and third features is as follows. For $a=1000$~m, 31\% and 19\% of the estimates at $\ell_{\rm samp}$ lie more than $3\lambda_{\min}(\bm{{\cal I}}_M)^{-1/2}$ from the true $\delta\bm p$, against 0.3\% for a Gaussian estimator with covariance $\bm{{\cal I}}_M^{-1}$. That is, the estimator often selects a distant offset at which the map predicts similar measurements, and the local bound does not describe this ambiguity. The fraction of such estimates is largest on short paths. For $a=1000$~m, it falls from 37\% and 29\% for $\ell_{\rm samp}<100$~m to 15\% and 7\% for $\ell_{\rm samp}\ge400$~m. By Equation~(\ref{eq:pathlength}), a large $\kappa$ predicts a short $\ell_{\rm samp}$. Accordingly, the locations with the largest $\kappa$ are the most exposed to ambiguity. For $a=1000$~m, $\rho_{\rm S}(\kappa,\ell_{\rm ML}/\ell_{\rm samp})$ equals $+0.46$ and $+0.36$.

We emphasise that $\kappa$ inherits the locality of Equation~(\ref{eq:worstcrlb}). On both maps, $\kappa$ ranks the distance that the estimator of Equation~(\ref{eq:ml}) requires more closely than $\|\nabla B\|$ for $a\le500$~m. It does not describe ambiguity at larger offsets.

\section{DISCUSSION}
\label{sec:discussion}

The cubic short-path law of Section~\ref{sec:cubic} applies to a scalar map of any differentiable field, e.g.\ terrain elevation, the gravity anomaly, or the magnetic anomaly. That is, Theorem~\ref{thm:cubic} uses only the gradient $\bm g_0$ and the Hessian $\bm H_0$ of the map $B$ at the start of the path $\bm x_0$, together with the scalar form of Equation~(\ref{eq:measurement}). From a physical point of view, the cubic coefficient $\kappa$ can be decomposed into a contribution from the curvature $k_{\rm c}$ of the contour and a contribution from the curvature $k_{\rm g}$ of the gradient line at $\bm x_0$, viz.\ $\kappa=\|\bm g_0\|^2(k_{\rm c}^2+k_{\rm g}^2)/12$, where $k_{\rm c}=\nabla\theta(\bm x_0)^\top\hat{\bm g}_{\perp,0}$, $k_{\rm g}=\nabla\theta(\bm x_0)^\top\hat{\bm g}_0$, and $\hat{\bm g}_0=\bm g_0/\|\bm g_0\|$. Specifically, the heading of Equation~(\ref{eq:optimalheading}) is parallel to $k_{\rm c}\hat{\bm g}_{\perp,0}+k_{\rm g}\hat{\bm g}_0$, i.e.\ it follows the contour where the contours curve, and the gradient where the gradient lines curve. A map with straight, parallel contours has $k_{\rm c}=k_{\rm g}=0$ and gives $\lambda_{\min}(\bm F_\gamma)=0$ on every path. For a given Hessian $\bm H_0$, the gradient enters $\kappa$ only through its direction. On \texttt{Renfrew\_395} and \texttt{Eastern\_395}, $\rho_{\rm S}(\|\nabla B\|,\|\nabla\theta\|)$ equals $-0.33$ and $-0.28$, respectively, indicating that the gradient axis tends to turn more slowly where the gradient is strong. The mechanism is specific to scalar measurements. A measurement of several field components with known orientation, e.g.\ a vector magnetometer or a gravity gradiometer, contributes $\bm J_j^\top\bm\Sigma^{-1}\bm J_j$ per sample, where $\bm J_j$ is the Jacobian of the measured components with respect to $\delta\bm p$ and $\bm\Sigma$ is their noise covariance. This matrix generically has rank two.

In practice, the measurement often carries an unknown offset, e.g.\ a barometric altitude error in terrain-referenced navigation or uncorrected diurnal variation in MagNav [idealisation (ii) of Section~\ref{sec:information}]. An offset whose change over the duration of a path is much smaller than $\sigma$ acts as a constant bias $b$ on every $z_j$. Such a bias removes the information carried by the mean gradient along the path. That is, setting $\bm\eta=b$ in Equation~(\ref{eq:schur}) and marginalising $b$ replaces $\bm F$ by the centred path information $\bm F_c=\int_0^L(\bm g-\bar{\bm g})(\bm g-\bar{\bm g})^\top\,{\rm d}s$, where $\bar{\bm g}$ is the path-mean gradient. Specifically, along a straight path with heading $\bm u$, $\bm g-\bar{\bm g}=\bm H_0\bm u\,(s-L/2)+O(L^2)$ has rank one at leading order, and $\lambda_{\min}(\bm F_c)$ generically grows as $L^5$. With a bias, neither Equation~(\ref{eq:optimalheading}) nor $\kappa$ carries over.

A presurveyed map has a finite spatial resolution, e.g.\ set by the spacing of the survey lines, the grid cell size, or upward continuation to the flight altitude. The cubic coefficient $\kappa$ is a property of the map at that resolution. That is, $\kappa$ uses the second derivatives of $B$, i.e.\ the Hessian $\bm H$. Specifically, each horizontal derivative multiplies the Fourier transform of $B$ by ${\rm i}\bm q$, where $\bm q$ is the horizontal wavevector, so $\bm H$ carries an additional factor of the wavenumber $\|\bm q\|$ relative to $\nabla B$. Smoothing and upward continuation attenuate the components with large $\|\bm q\|$ and lower $\kappa$ more than $\|\nabla B\|^2$. For example, on \texttt{Renfrew\_555}, the median ratio of $\kappa$ to its value on \texttt{Renfrew\_395} equals 0.387, against 0.579 for $\|\nabla B\|^2$. The median $\ell_{\rm samp}$ increases by a factor 1.38, consistent with $\ell_{\rm req}\propto\kappa^{-1/3}$ in Equation~(\ref{eq:pathlength}). The results in Section~\ref{sec:application} are specific to two aeromagnetic anomaly maps at three smoothing widths. Both $\ell_{\rm samp}$ and $\kappa$ follow from Equation~(\ref{eq:measurement}) on the smoothed map. The ranking in Section~\ref{sec:ranking} tests the leading term of Theorem~\ref{thm:cubic} against the finite-length bound of the same model, and Section~\ref{sec:estimator} tests that bound with a simulated estimator. Map error, inertial drift, and measurement bias lie outside the scope of both tests. The difference between a map and its smoothed version varies on lengths $\lesssim w$, comparable to the paths of Section~\ref{sec:application}. Equation~(\ref{eq:measurement}) omits this difference, which is an error correlated along the path. For white sensor noise of one-sided power spectral density $S_n$, sampled at the rate $f$ with the Nyquist bandwidth $f/2$, $\sigma^2=S_nf/2$, and Equation~(\ref{eq:pathlength}) gives $\ell_{\rm req}\propto(\Delta s\,\sigma^2)^{1/3}=(vS_n/2)^{1/3}$, i.e.\ $\ell_{\rm req}$ is independent of $f$.

\section{CONCLUSION}
\label{sec:conclusion}

In this paper, we formulate scalar-map localisation of a path with a constant horizontal position error $\delta\bm p$ in terms of the Fisher information accumulated along the path, and derive its short-path limit. The largest weakest-direction path information grows as $\kappa L^3$ [Equation~(\ref{eq:cubic})], where $\kappa$ depends only on the gradient and the Hessian of the map at the start of the path. A straight path along the heading $\bm u_0$ of Equation~(\ref{eq:optimalheading}) attains the leading term. Through the Cram\'er--Rao lower bound, $\kappa$ gives the distance scale $\ell_{\rm req}$ of Equation~(\ref{eq:pathlength}) at which the worst-direction standard deviation of $\delta\bm p$ reaches $\sigma_{\rm req}$, without a search over headings and path lengths. Because $\kappa$ requires only the map, it can be computed before a mission to screen map regions, to choose an initial heading, and to set the sensor noise needed to reach $\sigma_{\rm req}$ within a given distance.

In Section~\ref{sec:application}, on the \texttt{Renfrew\_395} and \texttt{Eastern\_395} aeromagnetic anomaly maps, the median ratio $\lambda_{\min}(\bm F_\gamma)/(\kappa L^3)$ along $+\bm u_0$ equals 0.996 and 0.998 at $L=30$~m, respectively. For $\sigma_{\rm req}=100$~m, the median ratio $\ell_{\rm samp}/\ell_{\rm req}$ equals 0.865 and 0.890. The correlation $\rho_{\rm S}(\kappa,\ell_{\rm samp})$ equals $-0.96$ and $-0.97$, against $-0.52$ and $-0.56$ for $\|\nabla B\|$. The heading $\pm\bm u_0$ gives a distance within 10\% of $\ell_{\rm samp}$ at 81.7\% and 87.5\% of locations. For a maximum-likelihood estimator that searches a square of half-width 300~m, the median ratio $\ell_{\rm ML}/\ell_{\rm samp}$ equals 1.11 and 1.06. For half-widths up to 500~m, $\rho_{\rm S}(\kappa,\ell_{\rm ML})$ is more negative than $\rho_{\rm S}(\|\nabla B\|,\ell_{\rm ML})$ on both maps.

There are at least two ways in which the present work can be extended. First, an unknown measurement bias removes the cubic term along every straight path (Section~\ref{sec:discussion}), and the optimal short path with a bias can be derived. Second, a test on real flight data subject to inertial drift as well as map error can quantify how closely $\kappa$ and $\bm u_0$ predict the distance and the heading that a navigation filter requires in flight.

\bibliographystyle{IEEEtran}
\bibliography{main}

\end{document}